\documentclass{article}
\usepackage{spconf,amsmath,graphicx,hyperref}
\usepackage{xspace}
\usepackage{enumitem}
\usepackage{amsmath}
\usepackage{algorithm}
\usepackage{algpseudocode} 
\usepackage{booktabs}
\usepackage{xcolor}
\usepackage{caption}
\usepackage{tabularx}
\usepackage[numbers,sort&compress]{natbib}

\newcommand{\tool}{\textsc{CryptoScope}\xspace}
\newcommand{\dataset}{\textsc{LLM-CLVA}\xspace}
\title{\tool: Utilizing Large Language Models for Automated Cryptographic Logic Vulnerability Detection}
\name{%
Zhihao Li$^{1}$, Zimo Ji$^{2}$, Tao Zheng$^{1}$, Hao Ren$^{1}$, Xiao Lan$^{1,*}$\thanks{*Corresponding Author: Xiao Lan}%
}

\address{%
$^{1}$Sichuan University, Chengdu, China\\
$^{2}$The Hong Kong University of Science and Technology, Hong Kong SAR, China
}

\begin{document}
%
\maketitle
\begin{abstract}
Cryptographic algorithms are fundamental to modern security, yet their implementations frequently harbor subtle logic flaws that are hard to detect. We introduce \tool, a novel framework for automated cryptographic vulnerability detection powered by Large Language Models (LLMs). \tool combines Chain-of-Thought (CoT) prompting with Retrieval-Augmented Generation (RAG), guided by a curated cryptographic knowledge base containing over 12,000 entries. We evaluate \tool on LLM-CLVA, a benchmark of 92 cases primarily derived from real-world CVE vulnerabilities, complemented by cryptographic challenges from major Capture The Flag (CTF) competitions and synthetic examples across 11 programming languages. \tool consistently improves performance over strong LLM baselines, boosting DeepSeek-V3 by 11.62\%, GPT-4o-mini by 20.28\%, and GLM-4-Flash by 28.69\%. Additionally, it identifies 9 previously undisclosed flaws in widely used open-source cryptographic projects.
\end{abstract}
\section{Introduction}

\label{sec:intro}

Cryptographic algorithms and protocols are fundamental to securing computer systems, offering confidentiality, integrity, and authentication based on strong mathematical foundations. However, translating these principles into correct implementations remains challenging and error-prone~\cite{anderson1993cryptosystems}. Developers must implement algorithms accurately, handle inputs properly, select parameters carefully, and optimize performance, with mistakes potentially compromising entire systems. Furthermore, insufficient cryptographic expertise and the increasing use of large language models (LLMs) like GPT-3.5~\cite{gpt3.5} for coding assistance may introduce subtle vulnerabilities. Such flaws in widely used cryptographic libraries can propagate to numerous dependent projects, as exemplified by the critical ECDSA bypass vulnerability (CVE-2022-21449)~\cite{cve2022} in Oracle Java SE and GraalVM, which allowed attackers to forge digital signatures and bypass authentication.

Existing automated detection efforts largely target cryptographic API misuse~\cite{egele2013empirical,kruger2019crysl,rahaman2019cryptoguard,xia2024exploring,baek2024cryptollm}. In contrast, only a few studies~\cite{wycheproof,aumasson2017automated,nilizadeh2019diffuzz,vranken2022cryptofuzz} have explored the automated detection of cryptographic logic flaws~\cite{anderson1993cryptosystems}, which often suffer from limited automation, strong language dependencies, and restricted generalizability.

To address this gap, we propose \tool, the first LLM-based framework for detecting cryptographic logic vulnerabilities. It initiates with a pre-detection step verifying algorithm correctness and employs few-shot learning~\cite{mann2020language} combined with Chain-of-Thought (CoT) prompting~\cite{wei2022chain} to guide the LLM in analyzing code parameters and logic. We further build a cryptographic knowledge base by extracting diverse, multi-source domain information and integrate relevant knowledge via Retrieval-Augmented Generation (RAG)~\cite{lewis2020retrieval} to enhance reasoning accuracy. Detection results are output in a structured, developer friendly format.

We evaluate \tool on \dataset, a 92-sample benchmark covering real-world vulnerabilities, Capture The Flag (CTF)~\cite{mcdaniel2016capture} challenges, and synthetic cases in 11 programming languages, using the LLM-as-a-Judge~\cite{zhou2022large} framework across six representative LLMs~\cite{deepseekv3,gpt4omini,glm4flash,qwenllm,gemini15flash,claude35haiku}. Deployed on 20 open-source projects, \tool discovered 9 previously unknown cryptographic flaws, demonstrating its practical effectiveness.

The contributions of this work can be summarized as follows. 
\begin{itemize}[itemsep=0pt, parsep=0pt, topsep=2pt, leftmargin=*]
  \item \textbf{Benchmark:} \dataset, comprising 92 multi-language cryptographic vulnerability samples with manual reports and comprehensive evaluation metrics.
  \item \textbf{Framework:} \tool, a language-agnostic LLM-based system leveraging CoT and RAG for cryptographic logic vulnerability detection without code execution.
  \item \textbf{Empirical validation:} Strong experimental gains across architectures, validated by ablations, real-world discoveries, and improved human analysis through knowledge augmentation.
\end{itemize}

\section{Related Work}
\label{sec:RW}

Automated detection of cryptographic logic vulnerabilities generally falls into two main categories: test vector–based validation and fuzzing. Project Wycheproof~\cite{wycheproof}, developed by Google, provides a comprehensive suite of curated test vectors targeting known issues in cryptographic algorithms. It includes over 80 test cases and has helped uncover more than 40 implementation bugs. However, its use across different programming languages requires custom parsers and test harnesses, which can limit portability and scalability.

Fuzzing-based approaches have also gained traction. DifFuzz~\cite{nilizadeh2019diffuzz} identifies side-channel vulnerabilities by generating inputs that maximize differences in resource consumption between program variants. CDF~\cite{aumasson2017automated} integrates fuzzing with stateless test vectors to explore known edge cases in cryptographic operations. Cryptofuzz~\cite{zhou2023clfuzz} employs differential testing across cryptographic libraries by comparing algorithm outputs to detect inconsistencies, and, with the help of sanitizers, can also reveal memory-related issues. However, many fuzzing-based techniques depend on triggering specific failure conditions and require manual inspection of anomalous behaviors to assess the underlying flaw.

\section{METHOD}
\label{sec:method}

\subsection{Construction of the \dataset Benchmark}
\label{subsec:Benchmark}

In our preliminary research, we developed a novel benchmark, \dataset (\textbf{LLM} for \textbf{C}ryptographic \textbf{L}ogic \textbf{V}ulnerability \textbf{A}nalysis), to address the absence of specialized benchmarks for evaluating LLMs in detecting cryptographic logic vulnerabilities. Our dataset comprises:
\begin{itemize}[itemsep=0pt, parsep=0pt, topsep=2pt, leftmargin=*]
    \item Code samples of cryptographic logic vulnerabilities from CVE entries (57\%).
    \item High-quality cryptographic challenges from major international CTF competitions (30\%).
    \item Artificially constructed algorithm implementations violating cryptographic standards (13\%).
\end{itemize}

For each code snippet in the dataset, we conducted manual auditing to summarize the cryptographic logic vulnerabilities present in the code, which served as the ground truth. To comprehensively assess model performance, we defined four evaluation metrics:

\begin{itemize}[itemsep=0pt, parsep=0pt, topsep=2pt, leftmargin=*]
    \item \textit{Credibility Score}: A composite metric assessing relevance, informativeness, and logical soundness of reasoning. It serves as the primary indicator of model performance.

    \item \textit{Cosine Similarity}: Measures semantic similarity between generated and reference reasoning via sentence embeddings (MiniLM-L6-v2), with scores in [0,1].

    \item \textit{Semantic Match Rate}: Assesses semantic consistency using LLM-as-a-Judge to determine alignment with the reference. Scores range from 0 to 1.

    \item \textit{Coverage Score}: Estimates the proportion of informative and relevant content in the output, judged by LLM-as-a-Judge.
\end{itemize}

\subsection{The Architecture of \tool}
\label{subsec:ARCH}

\begin{figure}[htbp]
    \centering
    \includegraphics[width=\linewidth]{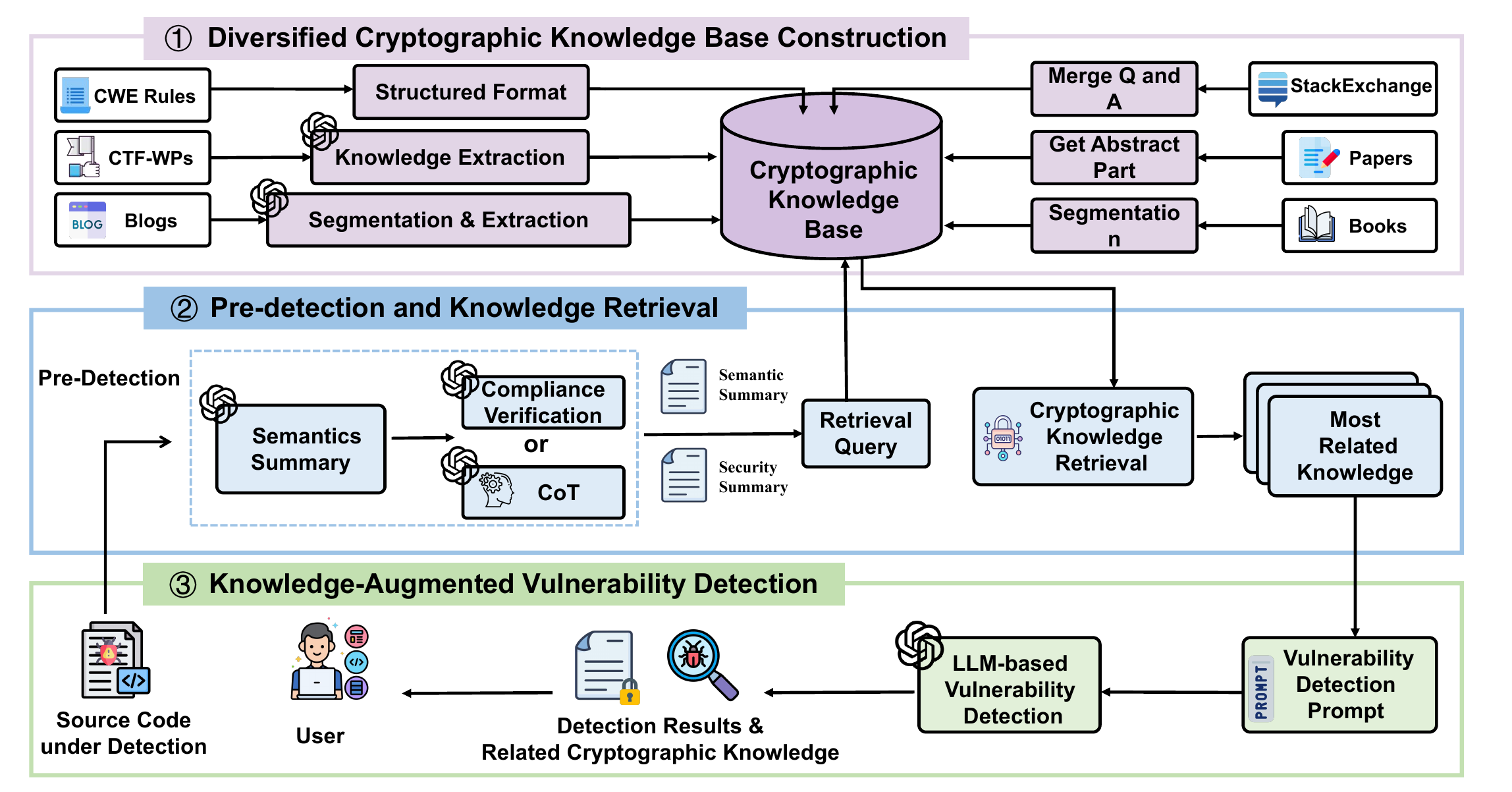}
    \captionsetup{skip=1.5pt}  
    \caption{Overview of \tool}
    \label{fig:overview}
    \vspace{-16pt}  
\end{figure}

In this work, we present a novel LLM-based cryptographic vulnerability detection framework \tool, the main idea of which is to leverage the semantic comprehension ability and the reasoning ability of LLM to simulate the process of cryptanalysts analyzing vulnerabilities. We summarize the process of manually analyzing cryptographic logic vulnerabilities into the following steps: understanding the semantics of the code, verifying its compliance with cryptographic algorithm standards, examining the code for potential vulnerabilities by referencing established vulnerability categories, and leveraging knowledge from vulnerability databases and best practices for cryptographic algorithm implementation to identify issues in the code. Accordingly, we applied this paradigm to the vulnerability detection model. Figure~\ref{fig:overview} shows the overview of our approach, which includes the following three phases.

\begin{itemize}[itemsep=0pt, parsep=0pt, topsep=2pt, leftmargin=*]
    \item \textit{Phase-1 Diversified Cryptographic Knowledge Base Construction}: Cryptographic knowledge is extracted from various unstructured documents via LLMs to construct the diversified cryptographic knowledge base.
    \item \textit{Phase-2 Pre-detection and 
    Knowledge Retrieval}: \tool summarizes the input code to extract its algorithmic and mathematical structure, then conducts a preliminary security analysis by either comparing it with cryptographic algorithm specifications or using few-shot CoT prompting. Both summaries are independently used to retrieve the most relevant knowledge block.
    \item \textit{Phase-3 Knowledge-Augmented Vulnerability Detection}: The LLM learns from the two retrieved knowledge blocks and conducts an in-depth analysis of code defects by integrating the pre-detection analytical process, ultimately deriving conclusions.
\end{itemize}

\vspace{-10pt}
\vspace{-4pt}
\subsection{Diversified Cryptographic Knowledge Base Construction}
\label{subsec:knowledge}

We construct a high-quality knowledge base of over 12{,}000 cryptography-related chunks, termed the \textit{diversified cryptographic knowledge base}, by crawling a wide range of open-source unstructured materials and applying large-model-assisted extraction and segmentation. Vectorized for efficient retrieval, this base supports downstream vulnerability detection.

\noindent\textbf{Data Sources.} The corpus integrates diverse cryptographic resources, summarized in Table~\ref{tab:kb-sources}.

\begin{table}[htbp]
    \centering
    \caption{Sources of the Cryptographic Knowledge Base}
    \label{tab:kb-sources}
    \renewcommand{\arraystretch}{1.1}
    \footnotesize  
    \begin{tabularx}{\linewidth}{@{}l>{\raggedright\arraybackslash}X@{}}
        \toprule
        \textbf{Source Type} & \textbf{Description} \\
        \midrule
        298 CTF Writeups & CTF crypto challenge writeups from top competitions. \\
        11 Cryptographic Blogs & Expert blogs on common cryptographic flaws. \\
        15 CWE Rules & CWE rules related to cryptographic vulnerabilities. \\
        3 Books~\cite{stamp2007applied,ferguson2011cryptography,menezes2018handbook} & Books on cryptographic implementation and security flaws. \\
        738 Research Abstracts & Abstracts of cutting‑edge cryptanalysis research. \\
        3909 StackExchange~\cite{cryptoSE} Posts & Practical cryptography Q\&A from StackExchange. \\
        \bottomrule
    \end{tabularx}
    \vspace{-6pt}
\end{table}

\vspace{4pt}
\noindent\textbf{Knowledge Extraction.}  
For CTF writeups, we use an LLM to extract fine-grained knowledge units per challenge. Blogs in Markdown format are manually segmented by third-level headers, then parsed into structured units by the LLM. Other sources are preprocessed through heuristic or fixed-size chunking.

\vspace{4pt}
\noindent\textbf{Embedding.}  
Knowledge units are stored in JSON Lines format. During vector index construction, all sources are embedded using cosine similarity. StackExchange questions serve as retrieval keys; question-answer pairs are returned. For other sources, each unit is used as both the key and content.

\subsection{Pre-detection}

The pre-detection phase has three components: \textit{Semantic Summary}, \textit{Compliance Verification}, and \textit{CoT-Based Reasoning}.

First, the LLM generates a semantic summary of the target code, emphasizing cryptographic logic, parameter sizes, and algebraic structures to aid understanding and support retrieval.

Next, compliance with standards is verified. We manually prepare reference documents for 42 common algorithms based on FIPS~\cite{nistfips}, covering logic flow, parameter limits, and security assumptions. The LLM checks conformity by analyzing parameter generation and encryption/decryption, simulating manual audits. Results form a retrieval index.

For non-standard algorithm code, few-shot CoT prompting guides the LLM to detect potential flaws by breaking down security goals—confidentiality, integrity, authentication—into concrete checks. Prompts focus on typical issues like input validation, primitive misuse, and error handling. Representative cases enable expert-level reasoning and generalization, enhancing accuracy and interpretability.

\begin{figure}[htbp]
    \centering
    \includegraphics[width=0.48\textwidth]{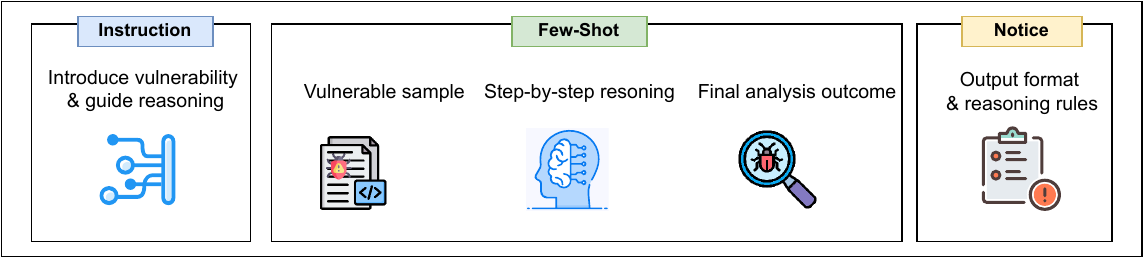}
    \captionsetup{skip=1.5pt} 
    \caption{CoT prompt structure.}
    \label{fig:CoT}
    \vspace{-7pt} 
\end{figure}

Figure~\ref{fig:CoT} shows the prompt format, consisting of three parts: Instruction (principle and reasoning steps), Example (code walkthrough), and Notice (output format and key reminders).

\vspace{-4pt}

To detect weak elliptic curves~\cite{novotney2010weak}, we integrate a remote SageMath~\cite{sage} execution environment. The LLM extracts curve parameters from the code, converts them into Sage-compatible syntax, and submits the code for remote execution. It then analyzes the returned results to assess potential vulnerabilities.

\subsection{RAG-based Vulnerability Analysis}
\label{subsec:rag}
Although CoT-based reasoning helps the LLM identify potential vulnerabilities, it is still prone to false positives, false negatives, and imprecise analyses. To mitigate this, we integrate a RAG mechanism that leverages external cryptographic knowledge to enhance and support the reasoning process.

Our retrieval strategy uses two signals: the semantic summary of the code and intermediate outputs from CoT-based reasoning. The former captures core cryptographic constructs and parameter attributes for matching structurally or semantically similar code fragments, while the latter highlights vulnerability logic paths that often correspond to known flaw patterns. These signals are embedded into vectors, and the top-$k$ relevant entries are retrieved using cosine similarity.

To avoid semantically irrelevant or misleading results, we apply a similarity threshold $\tau$ and retain only entries with $\mathrm{cos_sim} \geq \tau$. This process is detailed in Algorithm~\ref{alg:retrieval}. In subsequent experiments, we set the similarity threshold $\tau = 0.75$, as empirical results indicate this value achieves the best trade-off between relevance and precision.
\begin{algorithm}
\caption{Threshold-Based Cryptographic Knowledge Retrieval}
\label{alg:retrieval}
\footnotesize  
\begin{algorithmic}[1]
\Statex \hspace*{-1.8em} \textbf{Input:} Query $q$, number $k$, and threshold $\tau$
\Statex \hspace*{-1.8em} \textbf{Output:} Top-$k$ similar results $R$
\State $docs\_with\_scores \gets \texttt{similarity\_search}(q, k)$
\State Initialize result text $R \gets \emptyset$; counter $c \gets 1$
\For{each $(d, s)$ in $docs\_with\_scores$}
    \State $cos\_sim \gets 1 - s$
    \If{$cos\_sim \geq \tau$}
        \State Append content of $d$ to $R$ with index $c$
        \State $c \gets c + 1$
        \If{$d$ contains an ID}
            \State Extract ID and lookup full Q\&A from Crypto StackExchange
            \State Append question and answer to $R$
        \EndIf
    \EndIf
\EndFor
\State \Return $R$
\end{algorithmic}
\end{algorithm}

In the generation phase, the retrieved knowledge is combined with the original semantic and reasoning features and provided to the LLM. This design ensures that the model benefits from historical vulnerability patterns and domain-specific knowledge, while still retaining its internal reasoning capacity.
\vspace{-20pt}

\section{Evaluation Setup}

We evaluate the performance and practical value of \tool by addressing the following four research questions:

\begin{itemize}[itemsep=0pt, parsep=0pt, topsep=2pt, leftmargin=*]
    \item \textbf{RQ1: Compared to the baseline:} How does \tool perform compared to the baseline?
    \item \textbf{RQ2: Ablation study:} To what extent does each component of our framework contribute to the overall performance of vulnerability detection?
    \item \textbf{RQ3: Real-world evaluation:} How well does \tool detect cryptographic vulnerabilities in real-world cryptographic libraries?
\end{itemize}

\section{Results AND ANALYSES}

\subsection{RQ1: Effectiveness Compared to Baselines}

To evaluate the effectiveness of \tool, we compare it against vanilla LLMs on the \dataset benchmark across six representative LLMs using four metrics. The results in Table~\ref{tab:comparison_tool} demonstrate consistent improvements across all models.

\tool significantly enhances the performance of strong baselines such as DeepSeek-V3 and GPT-4o-mini. DeepSeek-V3 improves from 80.73 to 90.11 in Credibility Score and from 76.14\% to 83.04\% in Semantic Match Rate. GPT-4o-mini exhibits substantial gains in both credibility (+13.33) and coverage (+18.32\%). While the gains for Qwen-Plus and Gemini 1.5 Flash are more moderate, improvements are still observed across key metrics.

These findings indicate the generalizability of \tool: it systematically boosts cryptographic reasoning capabilities across diverse model backbones.

\begin{table}[ht]
\centering
\caption{Comparison of baseline LLMs and \tool on the \dataset benchmark.}
\label{tab:comparison_tool}
\resizebox{\columnwidth}{!}{
\begin{tabular}{lcccc}
\toprule
\textbf{Model} & \textbf{Credibility} & \textbf{Cosine Sim. (\%)} & \textbf{Semantic Match (\%)} & \textbf{Coverage (\%)} \\
\midrule
DeepSeek-V3 (Base) & 80.73 & 69.78 & 76.14 & 50.61 \\
DeepSeek-V3 (\tool) & \textbf{90.11} & \textbf{72.71} & \textbf{83.04} & \textbf{56.15} \\
\midrule
Qwen-Plus (Base) & 72.39 & 67.18 & 69.57 & 47.12 \\
Qwen-Plus (\tool) & \textbf{75.76} & 67.84 & 71.41 & 49.79 \\
\midrule
GPT-4o-mini (Base) & 65.74 & 61.45 & 61.96 & 36.38 \\
GPT-4o-mini (\tool) & \textbf{79.07} & \textbf{65.05} & \textbf{68.04} & \textbf{54.70} \\
\midrule
Gemini 1.5 Flash (Base) & 64.92 & 62.76 & 62.17 & 53.35 \\
Gemini 1.5 Flash (\tool) & \textbf{71.34} & \textbf{68.78} & 66.88 & 49.04 \\
\midrule
GLM-4-Flash (Base) & 53.93 & 60.40 & 48.21 & 28.62 \\
GLM-4-Flash (\tool) & \textbf{69.40} & \textbf{65.19} & \textbf{60.27} & \textbf{43.20} \\
\midrule
Claude 3 Haiku (Base) & 53.34 & 60.71 & 48.37 & 39.46 \\
Claude 3 Haiku (\tool) & \textbf{59.51} & \textbf{67.44} & 49.24 & 37.48 \\
\bottomrule
\end{tabular}
}
\end{table}
\vspace{-10pt}
\subsection{RQ2: Ablation Study}

We conduct an ablation study to evaluate the impact of two core components in \tool: the Pre-detection module (which employs Chain-of-Thought-based reasoning) and the Knowledge-Augmented Analysis module based on Retrieval-Augmented Generation (RAG). The experiments are performed on two representative large language models, DeepSeek-V3 and GLM-4-Flash, which serve as proxies for distinct categories of LLM architectures—DeepSeek-V3 representing open-source models optimized for multi-round reasoning tasks, and GLM-4-Flash exemplifying high-throughput models designed for efficient short-context inference. Table~\ref{tab:ablation} reports the resulting Credibility Scores under various configurations, highlighting the contribution of each component across different model capabilities.

\begin{table}[ht]
\centering
\small 
\setlength{\tabcolsep}{6pt} 
\caption{Ablation study results (Credibility Score).}
\label{tab:ablation}
\begin{tabular}{lcccc}
\toprule
\textbf{Model} & Baseline & Full & \textbf{w/o CoT} & \textbf{w/o RAG} \\
\midrule
DeepSeek-V3    & 80.73 & \textbf{90.11} & \textcolor{red}{83.02} & \textcolor{blue}{85.45} \\
GLM-4-Flash    & 53.93 & \textbf{69.40} & \textcolor{red}{65.32} & \textcolor{blue}{56.16} \\
\bottomrule
\end{tabular}
\end{table}
\vspace{-10pt}
\subsection{RQ3: Real-world Evaluation}

To assess practicality, we applied \tool (with DeepSeek-V3) to 20 real-world cryptographic codebases. The tool uncovered various logic-level flaws, such as improper ECDSA signature range checks, insecure padding in RSA, ECB-mode misuse, and weak key derivation practices.

Table~\ref{tab:github_vulns} presents representative cases. Notably, many issues had not been previously reported, confirming \tool's potential in real-world auditing.

\begin{table}[htbp]
\centering
\caption{Vulnerabilities discovered in open-source cryptographic projects.}
\label{tab:github_vulns}
\scriptsize
\setlength{\tabcolsep}{3.5pt} 
\begin{tabular}{|l|l|l|p{2.6cm}|}
\hline
\textbf{Project} & \textbf{Commit} & \textbf{File} & \textbf{Vulnerability} \\
\hline
goEncrypt & be7042 & rsacrypt.go & PKCS\#1 v1.5 misuse \\
\hline
cryptography & 5dc3c3 & controllers-ck.js & ECB mode, weak KDF \\
\hline
crypto-random-string & 25f893 & core.js & Modulo bias \\
\hline
nimcrypto & 4a0633 & pbkdf2.nim & Weak iteration count \\
\hline
generate-password & d11ddd & generate.js & Modulo bias \\
\hline
simple-crypto & 13559f & publickeysystem.py & Insecure RSA padding \\
\hline
ecurve & ee8a22 & curve.js & Incorrect square root algorithm \\
\hline
fastecdsa & 4617ef & \_ecdsa.c & Missing r/s range check allows signature bypass \\
\hline
crypto & 7112a2 & diffiehellman.py & Weak prime generation \\
\hline
\end{tabular}
\end{table}

\vspace{-10pt}
\section{Conclusion}

We introduced \tool, a novel framework for automated cryptographic logic vulnerability detection using LLMs. By combining CoT prompting and RAG with a curated knowledge base, CRYPTOSCOPE identifies complex flaws without code execution. On our \dataset benchmark, it consistently and significantly boosted the performance of various baseline models and discovered 9 undisclosed vulnerabilities in real-world projects. This work demonstrates that knowledge-augmented LLMs are a powerful, scalable, and language-agnostic tool for security auditing. Future work will enhance the knowledge base and reasoning capabilities.

\vfill\pagebreak




\begingroup
\setlength{\bibsep}{0pt plus 0.3ex}  
\small                                
\bibliographystyle{IEEEbib}
\bibliography{strings,refs}
\endgroup

\end{document}